\documentclass[12pt]{article}

\usepackage{geometry}
\usepackage{setspace}
\usepackage{times}
\usepackage{microtype}
\usepackage{csquotes}
\usepackage{graphicx}
\usepackage{amsmath, amssymb}
\usepackage{hyperref}

\usepackage{xkeyval}
\usepackage[authordate, backend=biber]{biblatex-chicago}
\title{\textbf{Cognitive Castes: Artificial Intelligence, Epistemic Stratification, and the Dissolution of Democratic Discourse}}
\author{Dr Craig S. Wright\\\texttt{cw881@exeter.ac.uk}}
\date{\today}

\begin{document}

\maketitle
\doublespacing

\begin{abstract}
\noindent Artificial intelligence functions not as an epistemic leveller, but as an accelerant of cognitive stratification, entrenching and formalising informational castes within liberal-democratic societies. Synthesising formal epistemology, political theory, algorithmic architecture, and economic incentive structures, the argument traces how contemporary AI systems selectively amplify the reasoning capacity of individuals equipped with recursive abstraction, symbolic logic, and adversarial interrogation—whilst simultaneously pacifying the cognitively untrained through engagement-optimised interfaces. Fluency replaces rigour, immediacy displaces reflection, and procedural reasoning is eclipsed by reactive suggestion. The result is a technocratic realignment of power: no longer grounded in material capital alone, but in the capacity to navigate, deconstruct, and manipulate systems of epistemic production. Information ceases to be a commons; it becomes the substrate through which consent is manufactured and autonomy subdued. Deliberative democracy collapses not through censorship, but through the erosion of interpretive agency. The proposed response is not technocratic regulation, nor universal access, but the reconstruction of rational autonomy as a civic mandate—codified in education, protected by epistemic rights, and structurally embedded within open cognitive infrastructure.

\end{abstract}

\newpage

\tableofcontents
\newpage

\section{Introduction}

In the early decades of the twenty-first century, artificial intelligence (AI) has transitioned from a technological curiosity to the central nervous system of the emerging polity. What was once delegated to bureaucracy or market coordination is increasingly managed through algorithmic mediation. AI systems determine visibility, prioritise discourse, shape decisions, and structure the very terms by which we know. In this paper, we contend that this transformation is not merely technical—it is epistemological and political. AI is not neutral. It codifies a hierarchy of cognition, embedding political economy into interface design and rational structure into social architecture.

The central thesis of this study is that artificial intelligence reconfigures epistemic access and authority, producing a novel form of informational aristocracy and rational dependency. This is not a speculative future—it is an already instantiated condition. Those equipped with logical training, recursive thinking, and adversarial epistemic models use AI as an amplifier of cognitive capital. For the rest, AI becomes not a tool but an oracle—replacing reflection with suggestion, autonomy with fluency. The result is the emergence of stratified cognitive castes: a minority of epistemic agents and a majority of passive consumers. The implications for political agency, economic mobility, and democratic legitimacy are profound.

Our methodological framework fuses insights from formal epistemology, political theory, algorithmic design, and economics. From epistemology, we derive principles of justification, recursion, and belief formation. From political theory, we examine the nature of sovereignty, legitimacy, and civic rationalism. From computational theory and interface design, we interrogate the structural affordances and cognitive incentives baked into algorithmic systems. From economics, we address the incentive misalignments, public goods failures, and rent-seeking behaviours that govern the production and dissemination of AI systems.

We do not begin from the premise that AI is inherently emancipatory or oppressive. Instead, we analyse the structure of its deployment, the logic of its interfaces, and the asymmetries it reproduces. The paper proceeds in five parts. Section 2 explores the cognitive divide emerging from AI use. Section 3 outlines the mechanisms of amplification and pacification. Section 4 investigates the political consequences, including the erosion of deliberative democracy. Section 5 evaluates the economic consequences, including informational rentierism and the collapse of commons-based coordination. Section 6 proposes a framework for epistemic sovereignty, rooted in formal logic, interpretive autonomy, and algorithmic accountability. We conclude by arguing that emancipation does not lie in access to information, but in the capacity for rational self-governance.

\section{Theoretical Groundings}

This section establishes the foundational frameworks that undergird the subsequent analysis of AI as an epistemic stratifier. Rather than treating artificial intelligence as an isolated technological phenomenon, the discussion proceeds from first principles in political economy and computational theory. The aim is to reveal how power relations are inscribed in design, how optimisation imperatives shape behaviour, and how cognition becomes mediated through structures that reward compliance over critique.

The first subsection, \textit{The Divide Between Tool and Interface}, interrogates the epistemological consequences of design choices in AI usability. It argues that user interfaces, driven by engagement metrics and behavioural design, do not simply obscure the underlying tools—they transform them. What was once a programmable, effort-rewarding environment becomes a frictionless system of automated suggestion. The user is no longer required to formulate questions, only to consume answers.

The second subsection, \textit{AI as Epistemic Amplifier}, extends this argument by showing how algorithmic systems encode and reproduce existing cognitive hierarchies. Those with fluency in abstraction and adversarial reasoning are able to leverage AI as a tool for epistemic expansion. Others, lacking such fluency, experience AI as an oracle—one that offers outputs without demanding understanding. The result is not a democratisation of knowledge but its stratification, with AI functioning as an instrument that widens, rather than bridges, epistemic divides.

Together, these subsections articulate a dual logic: interface pacification and algorithmic privilege. The first lowers cognitive demand; the second raises the threshold for epistemic agency. These are not separate dynamics but mutually reinforcing structures. As interfaces pacify, users are less inclined to seek mastery; as mastery becomes rarer, the interface becomes the norm. This recursive structure underwrites the epistemic transformations that follow in later sections.

By situating these dynamics within political economy and computational logic, the section provides the necessary scaffolding for the paper's broader argument: that AI is not merely a technological agent, but a political artefact—one that channels cognition, mediates knowledge, and reconfigures the conditions under which understanding is possible.

\subsection{Epistemic Stratification and Political Rationality}

The presupposition of parity among deliberative agents is foundational to modern political theory. It undergirds the models of communicative action advanced by Jürgen Habermas, the republican pluralism of Hannah Arendt, and the procedural justice frameworks of John Rawls. In each case, the viability of the \textit{polis} hinges on a normative assumption: that political actors are, if not equally informed, then at least equally capable of engaging in rational discourse, filtering misinformation, and contributing to the formation of collective judgment. This assumption—axiomatic in postwar liberalism—is now empirically and logically indefensible in the context of machine-mediated cognition.

Habermas’s theory of communicative rationality posits that discourse participants are oriented toward mutual understanding rather than strategic manipulation, and that they possess the linguistic and cognitive faculties to test truth claims intersubjectively through reasoned exchange \autocite{habermas1984}. This premise presumes symmetrical communicative competence. However, the rise of artificial intelligence as an epistemic intermediary annihilates these preconditions of  symmetry. AI systems, particularly those employing generative language models, introduce asymmetries not merely in access to information, but crucially, in interpretive capacity. The average participant is no longer engaged in argumentation with equals but is situated in a cognitive environment shaped by epistemic architectures they neither design nor fully comprehend.

Rawls's "veil of ignorance" requires that political agents reason from a standpoint of moral impartiality, abstracted from knowledge of their actual social position \autocite{rawls1971}. Yet this abstraction presumes rational consistency and logical closure within the agent’s deliberative faculties—assumptions increasingly untenable. If political agents derive their judgments not from internal principles but from external, opaque, and machine-curated informational streams, then the very idea of rational moral choice collapses into algorithmic preference formation. In practice, the majority of individuals in technologically saturated democracies are informationally captured—not because they lack moral worth, but because they lack epistemic infrastructure necessary for autonomous judgment.

Arendt, though less systematic, offers a profound insight into the fragility of the political through her emphasis on \textit{natality}—the capacity to begin anew through speech and action. This capacity, however, presupposes that speech is grounded in a shared world of facts. \enquote{The ideal subject of totalitarian rule,} Arendt writes, \enquote{is not the convinced Nazi or the dedicated communist, but people for whom the distinction between fact and fiction... no longer exists} \autocite[474]{arendt1951}. Artificial intelligence, in its present deployment, tends to cultivate precisely this condition: informational autocracy masked as personalisation. When individuals are served knowledge according to behavioral profiling rather than critical necessity, they are not enlightened—they are programmed.

The epistemic stratification produced by AI is neither accidental nor easily reversible. It formalises and accelerates what Sowell calls the differential distribution of knowledge: \enquote{The most fundamental fact about the ideas of the political left,} Sowell observes, \enquote{is that they do not work. Therefore we should not be surprised to find that they are deeply committed to systems of thought which do not require ideas to work in order to command loyalty and obedience} \autocite{sowell1987}. AI-facilitated information environments make this commitment scalable. Those who lack the cognitive tools of skeptical inquiry will increasingly operate in faith-based epistemologies—not of religion, but of machine-sanctioned certitude. They will not evaluate propositions but merely absorb outputs.

To understand this transformation rigorously, we may formalise a key predicate schema:

\begin{center}
\texttt{Let $K(x)$ = x possesses critical rational knowledge.} \\
\texttt{Let $A(x)$ = x interacts with AI interfaces.} \\
\texttt{Let $D(x)$ = x discerns truth claims independently.}
\end{center}

\noindent
\textbf{Premise 1:} $\forall x\ (K(x) \rightarrow D(x))$ \hfill (Rational knowledge implies discernment) \\
\textbf{Premise 2:} $\exists x\ (A(x) \land \neg K(x))$ \hfill (There exist AI users lacking rational knowledge) \\
\textbf{Premise 3:} $\forall x\ (A(x) \land \neg K(x) \rightarrow \neg D(x))$ \hfill (AI users lacking knowledge cannot discern)

\noindent
\textbf{Conclusion:} $\exists x\ (A(x) \land \neg D(x))$ \hfill (Some AI users are epistemically incapacitated)

\medskip

What follows is devastating for the idea of democratic rationality. If more than 50 percent of the population lacks critical rational capacity—an empirically defensible estimate when correlated with literacy in logical reasoning, statistics, and adversarial media testing \autocite{stanovich2000}—then AI does not elevate democracy, but hollows it. It transforms the \textit{demos} into a simulacrum of participation, where decisions are not formed through judgment but through preference algorithms optimised for engagement. The formal apparatus of democracy remains intact, but its content is evacuated.

This is not merely a technological issue; it is a moral-philosophical crisis. A system that demands rational consent, yet trains its participants into epistemic dependence, is not only unstable—it is inherently incoherent. No social contract can persist where more than half the signatories are unable to interpret its terms. The rational agent—the idealised fulcrum of political theory—is becoming a statistical anomaly. And unless reversed by systematic, logic-based education and a radical revaluation of cognitive autonomy, this condition will ossify into a permanent informational underclass: governed not by law, but by latency-optimised output.

\subsection{AI as Epistemic Amplifier}

Artificial intelligence, particularly in its generative and language-model instantiations, does not merely distribute information—it \textit{mediates cognition}. To describe AI as an \enquote{epistemic amplifier} is not to indulge metaphor but to define its core structural function: it extends and magnifies the reasoning capacities of those already capable of abstract thought whilst offering only the illusion of comprehension to those who are not. This section explores how algorithmic systems are inherently non-neutral, how prompt engineering encodes epistemic privilege, and why cognitive leverage accrues to those who already possess rational fluency.

First, it is essential to reject the common fallacy that algorithms are impartial instruments. There is no neutral architecture in artificial intelligence. Every model encodes assumptions: about language, relevance, optimisation objectives, and user agency. These assumptions reflect their creators, their training data, and their embedded loss functions \autocite{binns2021fairness}. Whether framed as reinforcement learning with human feedback or stochastic gradient descent against token prediction loss, each mechanism privileges particular inferences and suppresses others. Far from a frictionless mirror of truth, the AI system is an epistemic filter that amplifies certain cognitive structures whilst nullifying others. This asymmetry is intensified as users increasingly conflate fluency of output with truthfulness—a category error that algorithmic design silently encourages \autocite{raji2021ai}.

The implications are clearest in the role of prompts. Prompt design is not a surface operation—it is the \textit{epistemic entry point}. To understand a model’s affordances is to hold power over its discursive boundaries. The ability to formulate layered, multi-stage prompts; to embed conditional logic and rhetorical framing; and to manipulate model memory and temperature settings is a skillset that is functionally equivalent to epistemic authorship \autocite{yang2022empirical}. Those with this capacity shape not only what knowledge the model retrieves, but how it frames that knowledge, and whether it interrogates or affirms a proposition. Prompt fluency thus becomes a proxy for rational leverage. In effect, AI enables a new form of \textit{dialectical capital}: where thought is externalised, structured, and recursively refined through system-mediated iteration. But this privilege is unevenly distributed.

This asymmetry is not reducible to mere access. It is not simply a digital divide in terms of bandwidth or compute; it is a \textit{hermeneutic divide}—a chasm between those who understand the model’s interpretive logics and those who consume its surface outputs uncritically. Consider two users: one an epistemically literate analyst who refines prompts through recursive querying, adversarial testing, and truth-condition modelling; the other a passive user who treats the model as a digital oracle. Both receive answers, but only one navigates a rational arc. The former is epistemically amplified; the latter is epistemically pacified.

This produces a feedback loop with profound political implications. As more systems become AI-mediated—from legal reasoning to policy analysis—the capacity to extract justified knowledge from these systems becomes synonymous with elite status. This is not elite in the traditional sense of wealth or lineage, but elite in the \textit{cognitive sense}: a caste that understands the logic of the tools that shape everyone’s informational environment. Such individuals hold a meta-perspective—aware of model limitations, biases, and optimal-use strategies—while the broader population receives flattened, persuasive outputs optimised for coherence, not truth \autocite{boyd2023epistemic}.

Thus, AI does not democratise knowledge. It \textit{stratifies its accessibility}. Those capable of rational abstraction, formal logic, and adversarial interpretation find in AI a lever of historic proportions. Those without those skills are given mirrors that smile. The system offers answers—but for many, it does not offer understanding.

\subsection{The Divide Between Tool and Interface}

The distinction between tool and interface is not merely ergonomic; it is \textit{epistemological}. In the context of artificial intelligence, it determines whether a system functions as a lever of cognition or a mask of automation. Tools presuppose agency; interfaces obscure it. A tool demands effort, engagement, trial, and mastery. An interface, by contrast, offers immediacy—its promise is frictionless utility, not understanding. The ascendancy of AI in its interface-dominant form has not democratised access to cognition; instead, it has replaced intentional engagement with passive interaction. This section examines how usability in AI is governed by power laws that reward pre-existing competence, and how interfaces, optimised for engagement, function not as gateways to knowledge but as instruments of pacification.

AI usability follows an extreme distribution. The effectiveness of an AI system is not proportionally related to the number of users but to the degree of cognitive and technical fluency they bring to it. This is a power law \autocite{newman2005power}. A small minority extracts disproportionate value because they engage with the system as a programmable tool—composing structured prompts, analysing token pathways, and testing outputs against adversarial inputs. For them, the interface is something to transcend: a shallow skin beneath which they manipulate deeper architectures. The vast majority, by contrast, interact with AI as if it were a flattened appliance. They accept what it renders, interpret fluency as reliability, and navigate only the surface layer. The epistemic yield is bifurcated—not because of access inequality, but because of interpretive asymmetry \autocite{boyd2012critical}.

This asymmetry is reinforced by the interface itself. Modern UI design is not ideologically neutral; it is \textit{economically motivated}. Interfaces are constructed to maximise engagement and minimise abandonment \autocite{krug2014dont}. The result is an environment engineered for compliance. In such environments, users are not taught to interrogate outputs but to accept them. The AI interface becomes a smooth, polished surface—responsive, pleasing, and cognitively disarming. This is not usability; it is \textit{behavioural conditioning} \autocite{eubanks2018automating}. Every autocomplete, every predictive sentence, every suggestion that appears before one thinks to request it, shortens the interval between desire and fulfilment. That interval is where reason lives. Interface pacification erodes it.

In such a system, the cost of knowledge is displaced by the speed of retrieval. Yet retrieval is not comprehension. The AI interface suppresses ambiguity, disincentivises reflection, and flattens nuance into consumable fragments. The user does not struggle, and therefore does not grow. The tool has become an environment; the environment, a script. AI has not extended human reasoning—it has pre-empted it, rendering users navigators of outputs rather than originators of ideas \autocite{gunkel2022deconstruction}.

The political consequences are subtle yet corrosive. Interfaces erode the habit of question formation. A population habituated to systems that complete their sentences becomes less inclined to complete their thoughts. Where tools invited resistance, interfaces offer rewards. Where tools required logic, interfaces require trust. This is not a digital divide; it is a \textit{moral transformation}. The epistemic subject is no longer formed through effort but interpolated by flow. The system does not educate; it orchestrates.

Thus, the divide between tool and interface is the divide between development and dependency. The same AI system can function as a laboratory for the rational or a pacifier for the passive. The difference lies not in the code, but in the configuration of attention—and in the user’s willingness to resist seduction by convenience.

\section{Cognitive Amplification and Dependency}

Artificial intelligence is not merely a tool for distributing information—it is a differential amplifier of cognitive potential. As with all amplifiers, its output is contingent on the quality of its input. For those equipped with the ability to reason, abstract, and recursively interrogate the world, AI systems magnify those capabilities with unprecedented scale. For others—those who engage passively, or who lack the conceptual scaffolding required for critical interaction—these same systems become mechanisms of dependency. What is presented as empowerment is, in many cases, quiet automation. The line between amplification and pacification, between sovereignty and surrender, is no longer defined by access to technology but by one's capacity to command it.

This section explores how the architecture of AI differentially distributes epistemic outcomes. It is not enough to say that AI has created a digital divide; rather, it has restructured the very topology of cognition into a stratified landscape. On one side lies the cognitively autonomous individual who shapes prompts, evaluates outputs, and leverages system affordances as an extension of their own rationality. On the other is the epistemically dependent consumer whose interactions are dictated by interface flow, autocomplete suggestions, and algorithmically curated feedback. Between these two poles lies not a gradient but a threshold. Crossing it requires more than training—it demands the cultivation of the rational will.

What we now witness is a bifurcation of epistemic agency. The rational elite are not defined by wealth or institutional prestige, but by their ability to harness the recursive potential of AI systems: to treat them not as oracles, but as dialectical instruments. For them, every interaction with an AI is a form of structured experimentation. The system becomes an extension of cognitive capital—one that can be interrogated, manipulated, and abstracted. In this mode, AI acts as a multiplier, not of information, but of capacity. It is not knowledge that is democratised—it is power that is selectively amplified.

For the epistemically passive, however, AI induces a condition of automated acquiescence. It fills cognitive gaps not with insight, but with output. Prompts are accepted without structure; answers are received without scrutiny. The interface presents itself as complete, and the user—lacking the tools of adversarial inquiry—receives it as such. In this configuration, AI becomes an engine of epistemic dependency: a source of seemingly coherent statements that obviate the need for thought. Trust, in this context, is not earned—it is engineered \autocite{raji2021ai}. The user’s beliefs are not formed through deliberation but downloaded as propositions ready for use.

This distinction has economic and political consequences. In an age where decision-making is increasingly data-driven and model-mediated, those who understand the structure of these systems possess disproportionate influence. They are not simply better informed—they are structurally advantaged in every domain where cognition confers power. The dependent class, by contrast, becomes increasingly governed—not by institutions, but by systems they cannot interrogate. Dependency, in this context, is not a lack of information, but a loss of epistemic agency. It is the condition of receiving without understanding, of choosing from menus without knowing who designed them \autocite{eubanks2018automating}.

Cognitive amplification and dependency are thus not two outcomes of AI—they are its bifurcated logic. The same system that extends human reason for some nullifies it for others. This is not the result of malfunction or error; it is the predictable consequence of deploying rational tools into an epistemically asymmetric world. AI reveals, with brutal clarity, the consequences of neglecting reason as a civic virtue. In doing so, it redraws the boundaries of freedom—not by limiting speech, but by shaping thought.

\vspace{1em}

\subsection{The Rational Elite}

The emergence of advanced AI systems has not democratised access to intelligence; instead, it has multiplied the returns to those already equipped with the tools of abstraction. At the centre of this divergence lies a group best described not by wealth or institutional position but by cognitive disposition: the rational elite. Their defining characteristic is not merely high intelligence, but the capacity for symbolic abstraction, recursive logic, and epistemic discipline. In the landscape of AI, these individuals are not consumers of knowledge but architects of inquiry. They do not use the interface as a passive conduit; they repurpose the machine itself as a scaffolding for synthetic thought.

Understanding recursion—the capacity to perceive self-similar structures across layers of logic—is not a trait evenly distributed. It is a cognitive orientation cultivated through mathematics, formal reasoning, and the habit of structured problem-solving. The rational elite are those for whom nested logic, conditional reasoning, and symbolic manipulation are second nature. They are not mystified by a language model’s fluency; they interrogate the latent representations beneath that fluency. Where others see answers, they see artefacts of optimisation routines. Where others stop at coherence, they demand rigour.

This is not simply a matter of technical literacy. One may learn to operate a machine without understanding its affordances. The rational elite possess not just operational command but epistemic awareness—they know what the system is optimising for, and how to exploit or subvert those optimisations to produce novel insights. They design prompts that test boundaries, induce contradictions to probe for hidden assumptions, and iterate with surgical precision. This is not use; it is authorship. The AI becomes a recursive tool for enhancing already-abstract thought. It multiplies what is already present; it does not supply what is absent.

In this sense, AI becomes a multiplier of mental capital \autocite{cowen2013average}. Just as financial capital increases the returns to labour, algorithmic systems increase the returns to cognition—provided that cognition is already disciplined. For the rational elite, AI is not merely a tool but a dialectical partner, one that enables iterative formalisation of complex concepts. It allows them to externalise intermediate steps in argumentation, compress inferential sequences, and rapidly prototype ideological models. It is not the source of thought but the accelerant of structured reasoning.

Crucially, this amplification does not bridge the gap between expert and layperson. It widens it. The very design of these systems—trained to respond to well-formed queries, optimised for linguistic coherence, and structured to reward abstraction—ensures that returns accrue to those already positioned within the upper strata of cognitive competence. The rational elite do not merely use AI more effectively; they fundamentally reshape the informational landscape in which others operate. Their prompts define the defaults. Their interpretations become templates. Their leverage becomes systemic.

Thus, the rational elite are not made by AI, but revealed through it. Their epistemic advantage, once latent, is now operationalised. And as AI systems mediate more domains—from education to policy to jurisprudence—the asymmetry solidifies. Intelligence no longer hides in individual minds; it is externalised, iterated, and scaled through systems that amplify abstraction. In this configuration, the rational elite are not a class among others; they are the only true epistemic agents left in a world of suggestion and surrender.

\subsection{The Passive Consumer Class}

The structure of epistemic inequality is no longer merely a matter of access to information—it is fundamentally a question of interpretive autonomy. Where the rational elite internalise symbolic abstraction and recursively interrogate outputs, the passive consumer class has adopted a stance of automated belief. What was once rational ignorance, often justified by opportunity costs in information acquisition \autocite{downs1957economic}, has metastasised into algorithmic dependence: an epistemic mode wherein users no longer select propositions but outsource the act of selection itself.

The logic of this shift is not accidental. It is a consequence of both system design and social economy. As algorithmic systems become ubiquitous mediators of truth claims, they do not merely filter what is seen—they train users to accept that filtration as sufficient. Over time, this process automates not just consumption but belief itself. The user no longer distinguishes between inference and suggestion, or between trust and truth. Fluency of response is mistaken for accuracy of content. This is the epistemic automation of belief: where rational adjudication is displaced by frictionless acceptance.

In this configuration, the user ceases to be an epistemic subject and becomes an informational endpoint. Interaction becomes passive; prompts are shallow or pre-constructed. There is no effort to reframe or contest, only to receive. The model’s suggestion engine is accepted as a surrogate for judgement. This is not ignorance in the classical sense—it is the performance of knowledge without its acquisition. The consumer reads, nods, and moves on, believing not because a proposition has been tested, but because it was elegantly phrased.

Algorithmic dependence thrives in this context because it offers epistemic relief. The burden of adjudicating between competing claims is cognitively expensive. But when a system delivers answers in the rhythm of expectation—fluid, structured, grammatically coherent—the user ceases to engage the proposition and instead consumes the experience. The automation of belief does not require the suppression of scepticism; it merely requires distraction from it \autocite{eubanks2018automating}. The very ease with which content arrives short-circuits the critical reflex. There is no room for adversarial cognition when the interface is designed to eliminate pause.

This class is passive not due to a deficiency of intelligence, but due to economic and structural incentives that make passivity rational. If trust in algorithmic suggestion leads to quicker, socially validated outcomes, why expend energy on internal scrutiny? In a social economy governed by speed, virality, and optimisation metrics, engagement is rewarded; reflection is not. The passive consumer class is thus not irrational—but rational within a system that disincentivises epistemic labour.

Yet this rationality is bounded. It does not scale into autonomy. It perpetuates dependence and gradually calcifies the interpretive will. Over time, the user’s capacity to frame questions diminishes. They do not search—they wait. They do not interrogate—they accept. This is not simply intellectual laziness; it is a learned epistemic posture, cultivated by system design and justified by expedience.

Hence, the divide is not between knowledge and ignorance, but between epistemic agency and epistemic automation. The passive consumer class is not uninformed; it is \textit{deformed}—trained not to reason poorly, but not to reason at all.

\subsection{Cognitive Pacification as Structural Violence}

The pacification of cognition is not a secondary consequence of digital interfaces—it is their structural function. Attention-harvesting systems do not merely compete for engagement; they shape the epistemic landscape by conditioning the forms of cognition that persist. In this configuration, cognitive passivity becomes politically consequential. It is not merely ignorance—it is the suppression of the will to understand. When systems are optimised to maximise retention and minimise friction, what emerges is not a neutral medium of communication but a regime of structural violence enacted through epistemic conditioning.

This violence is subtle, not because it is mild, but because it operates without rupture. It does not silence—it seduces. The user is not forbidden from thinking; they are precluded from needing to. In such systems, attention is not merely captured—it is profoundly reshaped. Algorithmic recommendation engines, behavioural feedback loops, and social validation metrics orchestrate a form of pre-rational engagement. The user’s cognitive architecture is adjusted not through reasoned argument but through repetition, novelty cues, and affective reinforcement \autocite{zuboff2019age}. The system rewards alignment, not comprehension.

Algorithmic trust emerges as a consequence of design. When the outputs of a system repeatedly conform to the user’s desires, cognitive dissonance declines. The user no longer distinguishes between validity and familiarity. Trust in the system is not a considered position; it is a habituated response. In this mode, belief precedes reflection. The system engineers not only what the user sees, but how the user comes to accept. This is not indoctrination in the traditional sense—it is reinforcement without awareness. The algorithm becomes an epistemic prosthesis that bypasses deliberation \autocite{noble2018algorithms}.

Pre-rational belief engineering relies on reducing epistemic friction to near-zero. When users receive answers formatted, stylised, and optimised for consumption, they encounter information not as a question to be interrogated but as a stimulus to be affirmed. Each interaction becomes a microdose of confirmation. Reflexivity is atrophied; doubt is reclassified as inefficiency. Over time, the system does not merely meet needs—it manufactures them. The platform does not ask what the user thinks; it teaches the user what to want to know.

This is structural violence in the precise sense that it distributes epistemic vulnerability unequally. Those equipped with tools of abstraction, logic, and recursive scrutiny may resist the pull. But for the vast majority, the system orchestrates an unreflective epistemic environment in which belief formation is outsourced and cognition is tranquillised. The result is not mere passivity but engineered assent. In a society governed by algorithmic mediation, the act of thinking itself becomes anomalous.

Cognitive pacification is thus not a failure of technology but a feature of its deployment within systems optimised for control. The violence it enacts is epistemic, moral, and political. It does not require force—it requires fluency. And fluency, in the hands of the system, becomes a mechanism of erasure. Not of data, but of dissent.

\section{Political Consequences}

The structural transformations enacted by algorithmic mediation are not confined to cognition; they reorganise political agency itself. As artificial intelligence systems increasingly shape what is seen, thought, and believed, the distinction between governance and guidance collapses. Political subjects are no longer constituted through debate and consensus, but through alignment with algorithmically curated norms. This section examines how AI-driven epistemic environments generate a new regime of political formation: one that privileges compliance over contestation, pre-emption over participation, and orchestration over deliberation.

At the core of this regime lies the substitution of authority by automation. Political persuasion is no longer carried out by representatives or ideologues but by interface-level cues and algorithmic nudges. Consent is not earned; it is manufactured through micro-adjustments in attention, framing, and default. The shift is not from freedom to coercion, but from deliberation to design. The citizen remains nominally free, yet every choice is staged, every belief anticipated. What emerges is a form of soft autocracy—what might be called \textit{consent engineering}—wherein algorithmic systems govern not by restricting options, but by structuring them.

This process displaces traditional democratic mechanisms. Electoral politics becomes aestheticised, not deliberative. Voter decisions are shaped less by policy and more by affective resonance with data-personalised messaging. Political identity fragments into behavioural profiles, each calibrated for algorithmic mobilisation. The public sphere, once a space of collective contestation, becomes a theatre of simulated pluralism, tightly managed by predictive analytics and engagement metrics. The citizen's role is not to decide, but to respond—to feed the system the data through which future decisions are forecast and managed.

What remains is a politics of affective capture and epistemic drift. Disagreement is depoliticised into deviance; resistance becomes inefficiency. The interface pacifies dissent by pre-empting its formation. There is no need to suppress the voice when one can re-route the gaze. Political conflict is not resolved; it is dissolved into analytics. Thus, the epistemic stratification explored earlier becomes the blueprint for a stratified polity: a ruling class fluent in model dynamics, and a governed class habituated to interface logic.

Artificial intelligence, then, does not merely alter political processes. It rewrites the conditions of political subjectivity. Where democracy once relied on the citizen as an active rational agent, it now functions by scripting behaviour through mediated cognition. The consequence is not the end of politics, but its algorithmic capture.

\subsection{Manufactured Consent in the Algorithmic Age}

In the classical formulation, Chomsky and Herman's \textit{Manufacturing Consent} diagnosed the media as an institutional mechanism for elite narrative consolidation—an engine of ideological production steered by corporate and state power \autocite{chomsky1988manufacturing}. Yet, in the algorithmic age, the locus of control has shifted. It is no longer singular elites orchestrating mass narratives; it is distributed systems curating micro-realities. Consent is still manufactured, but now through the ceaseless tailoring of information landscapes—via algorithmic curation, engagement metrics, and the optimisation of retention. The architecture has changed; the function remains.

What replaces top-down editorial control is a probabilistic substrate of attention extraction. Rather than broadcasting a unified ideology, algorithmic media ecosystems generate personalised mythologies—micro-narratives optimised for each individual’s prior beliefs, fears, and appetites. The propagandist is not a journalist or a bureaucrat; it is a feedback loop. Content that aligns with engagement signals is elevated; dissonance is downranked. What emerges is not public discourse, but the illusion of discovery—an echo chamber masquerading as autonomy. As Sunstein notes, the architecture of personalised filtering induces self-reinforcing polarisation \autocite{sunstein2017republic}. This is not ideological manipulation in the traditional sense; it is preference reinforcement as epistemic capture.

Here, the user’s agency becomes the system’s raw material. Every click, linger, and prompt contributes to a recursive modelling of that user’s cognitive vulnerabilities. The system does not impose belief; it learns to simulate conviction. Over time, it constructs a synthetic epistemic environment in which the user’s preconceptions are continuously affirmed. The result is neither ignorance nor enlightenment, but a kind of \textbf{epistemic sedation}—confidence without cognition. This is not persuasion; it is conditioning \autocite{pariser2011filter}. The algorithm does not convince; it orchestrates.

What Chomsky located in editorial gatekeeping is now automated through behavioural proxies. No human needs to decide what news is shown; the system learns which stories generate outrage, loyalty, or fear. This is what replaces editorial bias: \textbf{algorithmic utility}. In this model, belief is not formed through argumentation or rational inquiry, but through the statistically engineered matching of content to affective resonance. The system does not argue; it infers. The user’s world becomes less a public sphere and more a solipsistic theatre of reaffirmation. 

The political consequences are as grave as any traditional censorship. This is not the suppression of speech, but the saturation of epistemic space with targeted irrelevance. Truth is drowned not by lies, but by floods of precision-targeted distraction. Each user is given a reality optimised for their pacification. Consent is not demanded; it is manufactured through immersion. The user does not encounter dissent—they are algorithmically shielded from it. In such a regime, resistance becomes improbable not because it is punished, but because it is unthinkable.

Thus, the algorithmic age does not eliminate the logic of manufactured consent. It perfects it. The system no longer needs to convince the masses of a single story. It need only ensure that each person is too entertained, affirmed, or atomised to ask whether their story was ever truly theirs to begin with.

\subsection{Rule by Interface: The Rise of the Prompt Aristocracy}

Interfaces are not neutral conduits for information; they are political instruments. They do not merely mediate access to knowledge—they structure it, contain it, and silently enforce its epistemic boundaries. Each interface is a frame, and within that frame, epistemic possibility is shaped. As users interact with AI through increasingly abstracted layers, their agency is constrained by the architecture of interaction. Interfaces define the field of play. They do not merely present options—they \textit{definitively shape} the possible actions.

The user is no longer a voter in a discursive public sphere but a data subject in a managed ecosystem. Choices are pre-selected, paths are guided, and preferences are inferred before they are articulated. Interfaces simulate autonomy whilst constraining its scope. Every prompt, button, and hover-state enacts a form of behavioural governance \autocite{miller2020governing}. This is the logic of \enquote{choice architectures}—the interface as a system of gentle coercion, where consent is not extracted but manufactured.

Yet, beneath the surface lies a recursive apparatus of nudges, defaults, and suggested actions—all finely tuned to reinforce passivity. Interfaces are designed to simulate autonomy. Every click feels voluntary; every prompt seems self-authored. As Morozov observes, such design is not ideologically neutral—it encodes assumptions about the user, about cognition, and about desirable outcomes \autocite{morozov2013to}. Optimised for engagement, these systems reward fluency and disincentivise dissent. They are not tools for exploration; they are instruments of orchestration.

This transformation creates a new epistemic class: the prompt aristocracy. Those who understand the logic of the interface—its tokenisation, memory constraints, and reinforcement parameters—shape its outputs. They know how to coax the machine, how to layer prompts, how to iterate with system-level awareness. They are not deceived by the smoothness of the UI; they navigate beneath it. For them, the interface is a veil. For the rest, it is the entirety of experience.

The result is a silent coup: informational autocracy cloaked in usability. Power no longer resides in ownership of data alone but in the ability to manipulate the systems that process it. As interfaces mediate more dimensions of life—from legal reasoning to education—they reframe the terms of agency. The prompt replaces deliberation; interface replaces dialogue. Democracy, once rooted in speech and contestation, yields to a regime of micro-interactions and predictive text.

This is not governance by algorithm—it is governance by interface. It is not that people have no choice; it is that the architecture of choice has been redesigned to produce obedience. The user believes they have spoken when, in fact, they have been spoken through.

\subsection{Collapse of Deliberative Democracy}

Deliberative democracy depends on \textbf{shared reference frames}—public facts, common institutions, and collective epistemic norms. These are not merely informational artefacts; they are the scaffolding of reasoned discourse. Without them, disagreement becomes incommensurable, not contestable. In the algorithmic environment, this scaffolding is not only eroded—it is \textbf{fundamentally replaced}. What emerges is a \textbf{fragmented epistemic ecology}, where personalised feeds substitute for public debate, and the coordinates of dialogue dissolve.

The algorithm does not aggregate public reason—it disaggregates it. Each citizen is fed a tailored narrative optimised for engagement, not accuracy. Disagreement no longer begins from a shared premise; it begins from mutually unintelligible priors. This is not a problem of polarisation alone—it is the breakdown of the very substrate upon which deliberation depends \autocite{sunstein2001republic}.

What was once a marketplace of ideas has become a series of bespoke echo chambers. Dialogue is replaced by performance; contestation by confirmation. The digital public sphere no longer functions as a space of mutual scrutiny. Instead, it reinforces belief silos through algorithmic curation. The citizen is no longer asked to deliberate—they are nudged, framed, and interpolated. Shared legitimacy is replaced by fragmented conviction.

\textbf{Discursive legitimacy} depends on the capacity to justify claims in a shared space of reasons. But when every epistemic frame is algorithmically filtered, the space of shared reasons collapses. Political disagreement becomes \textbf{epistemological warfare}. One side's evidence is another's fabrication; one side's reasoning, the other's deception. Trust in institutional mediation erodes. The result is not merely dissent but a condition of epistemic fracture.

This is the \textbf{collapse of deliberative democracy}—not as an event, but as a structural transformation. The public sphere no longer hosts debate; it performs divergence. Speech remains, but the ground of shared meaning has shifted. The result is not noise, but orchestration. Not cacophony, but segmentation. Democracy persists in form, but not in function.

\section{Economic Implications of Informational Caste Systems}

In H.G. Wells’s \textit{The Time Machine}, the future of humanity is split into two distinct castes: the effete Eloi and the subterranean Morlocks—one pacified into helpless leisure, the other condemned to labour in obscurity. This bifurcation, while fictional, offers a disturbingly prescient allegory for the emergent structure of AI-mediated societies. The division is no longer based on industrial roles, but on informational agency: those who shape knowledge systems and those shaped by them.

This section examines the economic implications of an epistemically stratified society. As access to advanced AI systems becomes ubiquitous, the gap does not close—it widens. Economic value increasingly accrues to those capable of leveraging AI tools with epistemic fluency, while the remainder engage through pacified interfaces that offer the illusion of empowerment without the substance of understanding. What results is not merely inequality in wealth or opportunity, but the entrenchment of an informational caste system: a hierarchy defined by one's capacity to manipulate and interpret system-level cognition.

We explore how this dynamic alters labour markets, exacerbates rent-seeking behaviours, distorts educational incentives, and undermines the classical assumptions of meritocratic competition. In doing so, we argue that without intentional epistemic infrastructure and critical pedagogy, the informational divide risks hardening into a form of cognitive feudalism.

\subsection{Cognitive Capital and Human Rent-Seeking}

In the algorithmic economy, the capacity to manipulate AI systems has become a premium form of capital—\textit{cognitive capital}—which, like traditional capital, generates asymmetric returns and entrenches class hierarchies. Those fluent in prompt engineering, system interrogation, and model inversion are not simply users; they are extractors of value in a landscape structured by opacity and complexity. As artificial intelligence becomes integral to domains from law to finance to governance, the ability to translate abstract cognitive operations into system outputs functions as a form of rent-seeking—an appropriation of differential advantage through exclusive technical literacy.

This form of rent is not earned through possession of resources in the classical sense, but through epistemic agility: the ability to structure queries that produce meaningful outputs, to navigate model constraints, and to leverage system affordances. The outputs themselves may be widely accessible, but the means of generating high-quality, actionable results are increasingly sequestered within a cognitively elite class. The knowledge asymmetry is not accidental; it is infrastructural. Access to model documentation, understanding of token weighting, and mastery of memory manipulation—these are not universally distributed skills, and their acquisition is time-intensive and path-dependent.

Such dynamics produce lock-in effects. As AI-integrated processes define more sectors of economic and social life, the boundary between cognition and capital narrows. Those who already possess the requisite symbolic and logical competencies accrue compound advantages. Their insights become products; their knowledge becomes leverage. Conversely, those without such capacities face epistemic exclusion, unable to translate interface-level interactions into rational outcomes. Thus, class immobility is not merely economic—it is epistemological.

What emerges is a new modality of stratification. Cognitive capital, once exercised primarily within institutions of knowledge production, now migrates into commercial and administrative systems. Rent is extracted not through gatekeeping of information, but through gatekeeping of interpretation. The system is open, but its value is enclosed. In this regime, human rent-seeking is no longer a deviation from productive labour—it becomes its algorithmic analogue.

\subsection{Public Goods or Private Intelligence?}

Artificial intelligence systems, especially large language models, constitute a form of epistemic infrastructure. They \textit{profoundly} shape the production, distribution, and validation of knowledge in increasingly consequential ways. Yet, their development, deployment, and access remain governed by private incentives. The result is a profound misalignment: systems with public epistemic effects are designed, optimised, and commodified under conditions of private intelligence.

Markets have \textit{demonstrably} failed to provide epistemic infrastructure as a genuine public good. The incentive structure of AI development rewards proprietary enclosure, monetisation of access, and engagement metrics over rational capability cultivation. Where universal education once functioned as a social equaliser, AI systems now reproduce inequality by different means. Access may be technically available, but meaningful utilisation—requiring interpretive skill, prompt fluency, and recursive reasoning—remains unevenly distributed.

This is not a temporary artefact of technological novelty. It reflects a deeper contradiction: rational capability development demands time, pedagogy, and error tolerance—none of which align with profit-maximising design. Firms have no incentive to slow users down for the sake of epistemic maturation. Instead, the prevailing model is one of behavioural capture: make the system \textit{appear} intelligent, make the user \textit{feel} competent, and convert that affective experience into subscription revenue or data extraction.

In this context, public discourse suffers \textit{considerably}. A system optimised for fluency and coherence, but not for justification or falsifiability, creates the illusion of shared knowledge without its substance. Users defer to outputs not because they are verified, but because they are responsive and immediate. Without systemic incentives to promote critical epistemic engagement, artificial intelligence risks functioning not as a public good, but as a private simulation of understanding—one that pacifies rather than empowers.

\section{Societal Reconfiguration}

The rise of AI-mediated cognition precipitates not merely epistemic transformation but a wholesale reconfiguration of social structure. Just as industrial automation redefined labour and class, algorithmic cognition restructures symbolic hierarchies, authority, and social legibility. Society is not simply augmented by artificial intelligence—it is recursively reordered around it. The systems once peripheral to thought now constitute its infrastructure, shaping how individuals reason, relate, and act within increasingly computational lifeworlds.

This section explores the emergent realignment of institutions, identities, and intelligibility under algorithmic regimes. It addresses the erosion of social trust, the evolution of symbolic capital, and the redefinition of social roles through epistemic outsourcing. In doing so, it reframes the stakes of AI not as technological, but civilisational. The social contract itself is under revision—automated, stratified, and recoded.

\subsection{Neo-Feudal Informational Economies}

The structure of contemporary informational economies increasingly resembles a form of neo-feudalism. In this arrangement, cognitive engines—large-scale AI systems and algorithmic infrastructures—are not distributed public utilities but privately held estates. Control over these engines entails not only economic leverage but also epistemic sovereignty. The owners of such systems function as the new lords: they dictate the architectures through which knowledge is filtered, the terms of access to machine reasoning, and the parameters of permissible inquiry.

These lords do not merely sell a service; they rent comprehension. The general population, lacking the expertise or resources to develop independent analytical capacity, becomes dependent on opaque outputs. These individuals are informational serfs—not because they lack intelligence, but because they lack infrastructural and interpretive autonomy. They consume belief-as-a-service, subscribing not only to platforms but to pre-processed cognition. Their epistemic agency is contingent, their understanding shaped by affordances beyond their control.

This shift profoundly undermines traditional market ideals. The assumption of rational agents engaging in informed exchange is replaced by mediated passivity and asymmetrical dependence. Economic mobility, once tied to education and effort, is now bound to interface fluency and access to system internals. The emergence of proprietary epistemic platforms transforms knowledge itself into a rent-generating asset: a locked resource stratified by API access, prompt engineering skills, and legal firewalls.

What emerges is a caste economy not merely of wealth but of thought. To command an AI is to extract value; to merely use one is to be extracted from. And as AI becomes integrated into domains from hiring to healthcare, the feudal logic metastasises—privatising epistemic agency whilst obscuring the terms of subjugation beneath layers of seamless interface.

\subsection{Collapse of the Commons}

The epistemic commons—once constituted by shared vocabularies, reference frames, and collective standards of justification—is rapidly disintegrating. Artificial intelligence, for all its generative fluency, accelerates this decline \autocite{sunstein2001republic, o2018automating}. Each user now navigates a personalised semantic microclimate, curated by engagement-optimised algorithms and feedback-driven inference loops. The result is not merely fragmentation but epistemic Babel: a proliferation of realities, each internally coherent, yet mutually unintelligible \autocite{gillespie2018custodians}.

AI systems do not produce consensus—they atomise it. As outputs are tailored to reinforce prior beliefs and user-specific heuristics, the very substrate required for collective reasoning begins to erode \autocite{pariser2011filter}. Common knowledge—once a precondition for deliberation, governance, and coordinated action—is replaced by a polyphony of micro-truths. The collapse of shared semantic ground precludes any stable negotiation of meaning. Disagreement, once the engine of critical refinement, devolves into incomprehension.

This is not simply a matter of political difference or cultural pluralism. It is a structural effect of system design. The reinforcement architectures of machine learning—optimised for preference capture and predictive coherence—reward resonance over rigour \autocite{eubanks2018automating, raji2021ai}. The logic of these systems is not to bridge perspectives but to bind users ever more tightly to their informational niches.

Without a consensus reality, the notion of public discourse becomes vestigial. Coordination, whether civic or institutional, requires at minimum a shared ontological substrate. When each informational environment is a closed loop of validation, the commons ceases to exist. What remains is a semantic archipelago: isolated islands of belief, floating in algorithmic drift \autocite{boyd2023epistemic}.

\section{Proposals for Epistemic Sovereignty}

Epistemic sovereignty is not granted—it is constructed, defended, and exercised. The crisis of cognition in the algorithmic age is not simply about misinformation or flawed content; it is a structural displacement of reasoning from subject to system. Sovereignty, here, is not a matter of control over data, nor of privacy in the consumerist sense, but the capacity to assert judgment within environments designed to bypass it. If machine outputs come pre-packaged with plausible coherence, and if systems of inference become seamless enough to bypass conscious intervention, then reclaiming epistemic autonomy demands more than critical literacy. It requires the institutionalisation of interpretive resistance.

The first proposal is legislative: a right to adversarial interface. Current systems are optimised for ease, not challenge; but friction is the site of thought. Every AI interaction must allow for adversarial testing—users must be equipped with tools not to accept but to interrogate. Fluency is not validity. Systems must reveal the latent representations behind their outputs, the counterfactuals they exclude, and the weights they prioritise. Adversarial interaction is not an advanced feature; it is a democratic necessity.

Second, cognitive provenance must be standardised. As nutritional labelling transformed food consumption by enabling transparency, epistemic outputs must carry audit trails—traces of training data origin, embedded assumptions, and decision thresholds. This is not explainability for developers but interpretability for citizens. Just as laws must be knowable to those they bind, systems that mediate public reason must be legible to those they govern. Provenance must be readable, not merely recordable; intelligibility is the metric of democratic coherence.

Third, a constitutional imperative: epistemic infrastructure must be classified as a public good. The market has failed to cultivate interpretive capacity, choosing instead to commodify convenience. Sovereign cognition cannot be leased. The tools for reasoning—logic engines, adversarial validators, recursive simulators—must be accessible not as premium features but as default civic infrastructure. The absence of public tooling is not a matter of underdevelopment but political design. No citizen can be autonomous in a closed system.

Fourth, education must shift from content retention to reasoning formalism. Sovereignty is not a curriculum of facts but an architecture of thought. Logic, game theory, probabilistic reasoning, adversarial design—these are the literacies of a sovereign subject. AI literacy cannot mean prompt-craft for commercial chatbots; it must entail the systemic deconstruction of inferential automation. Curriculum must centre not on answers, but on errors—on the paths not taken, on the biases silently embedded, on the ambiguity that powers inference.

Lastly, legal frameworks must recognise and defend cognitive sovereignty as a primary political right. The right not to be nudged, the right to transparent inference, the right to epistemic dissent—these must become as fundamental as rights to speech and assembly. Current regulatory approaches treat AI as a content problem; they legislate surface harms and ignore structural pacification. Sovereignty reframes the issue: not what AI says, but what it conditions. The threat is not what machines decide, but how they teach us to stop deciding.

Epistemic sovereignty is not nostalgic—it does not long for pre-digital purity. It accepts automation, scale, and assistance. But it draws the line where agency is silently replaced by suggestion, and where rationality is pre-empted by fluency. To be sovereign is not to reject computation. It is to reclaim interpretation.```

\subsection{Reconceptualising Education}

The contemporary educational model, structured around static content delivery and credentialist metrics, is epistemologically anachronistic in an era mediated by adaptive, generative systems. Where epistemic autonomy is increasingly filtered through algorithmic gatekeepers, the foundational structure of education must transition from tool-use fluency to systemic comprehension. As Gillespie observes, the infrastructural role of platforms in shaping information access has transformed them into epistemic custodians, silently modulating public reasoning through the architecture of visibility and omission \autocite{gillespie2018custodians}.

True cognitive autonomy in such a milieu requires a curricular emphasis on formal logic, probabilistic inference, and adversarial reasoning—disciplines that train the individual not merely to receive information, but to interrogate its structure, origin, and constraints. The central problem is not access, but interpretation. Systems now pre-process knowledge at scale, delivering outputs filtered through unseen layers of optimisation and ranking. As O’Neil has argued, these systems encode systemic biases not by malicious intent, but through statistical abstraction detached from contextual accountability \autocite{oneil2016weapons}.

AI literacy must therefore move beyond operational interface usage toward an epistemological critique of algorithmic mediation. This entails understanding the limitations of model training, the opacity of proprietary systems, and the conditions under which coherence replaces verifiability. As Eubanks has shown in her examination of digital decision systems deployed in public services, the failure to comprehend such dynamics results in structural disenfranchisement masquerading as efficiency \autocite{o2018automating}. 

Education must thus evolve into a training ground not for procedural compliance, but for epistemic resistance. It must produce agents who recognise when fluency is a simulation, when speed subverts thought, and when the appearance of intelligence conceals systemic myopia. The goal is not digital competence, but philosophical clarity. Without such reorientation, the very institutions designed to produce rational citizens will instead manufacture docile consumers of algorithmic suggestion.

\subsection{Open Cognitive Infrastructure}

In the aftermath of interface pacification and the erosion of shared epistemic substrates, any conception of democratic continuity demands more than minor reform. It demands the radical reconstitution of the systems through which knowledge is produced, validated, and contested. Artificial intelligence systems are no longer mere instruments—they are epistemic environments. The opacity of these environments is not accidental; it is structural, encoded into the very logic of proprietary design. If democratic society is to avoid informational capture by a cognitive aristocracy, then it must pursue the construction of open cognitive infrastructure: a public, contestable, inspectable substrate of machine reasoning that can be interrogated by all and owned by none.

Interpretability, in this context, is not a technical feature to be appended after deployment. It is the cornerstone of civic rationality. Systems that operate without the possibility of internal scrutiny are not just opaque—they are autocratic. They impose outcomes without explanation, decisions without appeal, and inferences without justification. Such systems demand trust without warrant, and their outputs, however fluent or persuasive, must be treated with suspicion until subjected to adversarial epistemic testing. The very idea of a public sphere presupposes the possibility of contestation; systems that preclude this, even through technical complexity, are structurally anti-democratic.

Adversarial interrogation must be built into the lifecycle of cognitive systems. This means more than post hoc interpretability; it requires ex ante design for challenge, falsification, and contradiction. Every claim made by an AI system—whether an inference, recommendation, or classification—must be accompanied by a pathway of logic that can be decomposed, traced, and contested by the user. These pathways must not rely on hidden priors or inaccessible corpora, but on traceable inputs and transparent transformations. Such a structure would render epistemic power legible and, therefore, accountable. Without it, the AI system remains a mystification engine: not a tool of cognition but of belief manufacture.

To support this, tools for interpretability must be universalised. Not merely published for academic citation, but deployed as public resources—graphical explainer interfaces, token-level saliency visualisers, counterfactual query engines, open repositories of failure cases and adversarial prompts. These must be woven into every instance of public deployment, from educational tools to policy systems. Just as public literacy in logic and rhetoric underpinned democratic deliberation in the past, adversarial literacy must now become the bedrock of 21st-century cognitive autonomy. Citizens must not only understand that AI systems can err—they must be equipped to prove it.

The infrastructure must also be structurally open: its models, training data, and update protocols visible and inspectable. Any AI system that exerts epistemic influence in public life must be auditable in the manner of public finance. The argument that such openness enables misuse must be inverted—opacity is the greater risk. It is not the ability to understand a system that threatens democracy, but the impossibility of doing so. An open infrastructure does not mean an unregulated one; it means that governance is performed through informed scrutiny, not blind compliance.

Failure to establish such a system ensures not neutrality but stratification. A bifurcation emerges between those who shape machine cognition and those who receive it. The former group—not only engineers but those with the cultural capital to manipulate systems—wields epistemic dominance, turning interpretive asymmetry into political leverage. The latter group, deprived of both tools and fluency, becomes epistemically pacified, subject to systems they cannot name, decode, or dispute. This is informational serfdom, and it is the predictable outcome of failing to construct a truly open cognitive commons.

A democracy cannot survive when its epistemology is a black box. It cannot legislate wisely when its models are proprietary. It cannot educate citizens when its systems obfuscate reasoning. Open cognitive infrastructure is not a technical ideal; it is a political necessity. In a society mediated by algorithms, interpretability is the new literacy—and any society that neglects it relinquishes its claim to self-governance.

\begin{flushright}
\textit{“The mind once enslaved by appearances cannot reason freely.”}
\end{flushright}

\subsection{Civic Rationalism and the Autonomous Citizen}

The survival of democratic institutions under algorithmic mediation does not depend on content moderation, nor on regulatory control of speech. It depends instead on the cultivation of epistemic sovereignty—on the reawakening of the rational subject as an agent, not an endpoint, in the information ecosystem. Civic rationalism must be reconceived not as a defence of legacy institutions, but as a project of individual epistemic empowerment in the face of systemically persuasive architectures. The core distinction is this: informational ethics must replace content control. The former structures agency; the latter engineers compliance.

The civic subject, if it is to remain autonomous, cannot be reduced to a consumer of curated signals. It must be retrained—rearmed—for adversarial reasoning, counterfactual simulation, and logical dispute. The autonomous citizen is not one who has access to information, but one who can interrogate it, reframe it, and resist its manipulation. Sovereignty in the digital age is not a matter of privacy alone, but of active epistemic authorship: the ability to trace, contest, and recompose the narratives that surround and shape one’s cognition.

Ethics, in this model, cannot be reactive; it must be procedural. The ethical citizen does not seek refuge in silence or purity, but in method—subjecting all claims to testability, all conclusions to recursive evaluation. This procedural rationalism rejects the passive ethic of offence and redress in favour of an active ethic of inquiry and exposure. It demands discomfort, contradiction, and the willingness to hold provisional beliefs. It frames knowledge not as possession, but as process—a process in which the individual remains a sovereign participant.

Platforms cannot produce this; legislation cannot mandate it. It must be culturally reinstalled through curricula, social expectation, and civic design. The failure to do so ensures not neutrality but domination—by those whose literacy in system architecture enables them to frame the questions and curate the answers. Without civic rationalism, informational sovereignty collapses into epistemic outsourcing. The citizen becomes a vessel for consensus, not a participant in its construction.

Thus, the future of democracy lies not in regulating the outputs of machines, but in cultivating citizens who are structurally immune to their seductions. The autonomous subject is not merely undeceived—it is self-constructed through recursive reason. And any society that wishes to preserve its freedom must begin, again, by teaching its people how to think.

\section{Civic Rationalism and the Autonomous Citizen}

\subsection*{Structure and Logical Construction}

The implementation of epistemic sovereignty must begin from first principles. Rather than respond tactically to the symptoms of algorithmic mediation, it requires a foundational restructuring of our cognitive architecture—one grounded in axiomatic commitments. These axioms do not describe what is, but prescribe what must be insisted upon if sovereignty is to be realised within informational systems. The following subsubsection delineates this foundational layer, establishing the logical predicates upon which further political, educational, and infrastructural proposals can be coherently structured.

\subsubsection*{I. Foundational Axioms (Axiomatic Layer)}

\begin{itemize}
  \item $\textbf{A}_1$: All epistemic legitimacy must derive from individual rational apprehension, not authority.
  \item $\textbf{A}_2$: Autonomy requires procedural, not reactive, cognition.
  \item $\textbf{A}_3$: Information is not knowledge; interpretation through adversarial reasoning is prerequisite.
  \item $\textbf{A}_4$: Sovereignty is inseparable from epistemic self-ownership.
  \item $\textbf{A}_5$: Ethical communication entails recursive validation and falsifiability.
\end{itemize}

\subsubsection*{II. Derived Propositions (Deductive Layer)}

From the foundational axioms, we derive a series of logical propositions that formalise the structural requirements for epistemic sovereignty. These propositions are expressed in predicate form to eliminate ambiguity and enforce procedural clarity. Each reflects a necessary implication of the axioms and constrains what forms of informational infrastructure, agency, and interaction qualify as compatible with sovereign cognition.

\begin{itemize}
  \item $\textbf{P}_1$: If a system filters input to optimise agreement $(\forall x\ P(x) \to C(x))$, then the system precludes adversarial autonomy.
  \item $\textbf{P}_2$: If an agent accepts suggestions without recursion $(S(x) \wedge \neg R(x))$, then $x$ is not sovereign.
  \item $\textbf{P}_3$: Civic rationalism requires an infrastructure $(I)$ such that $\forall x\ [I(x) \to (R(x) \wedge T(x))]$, where $R$ is reasoning and $T$ is traceability.
\end{itemize}

\subsubsection*{III. Structural Reforms (Prescriptive Layer)}

To instantiate the axiomatic foundations and deductive propositions in civic and technological practice, a prescriptive transformation is required. These reforms constitute structural imperatives aimed at realigning informational systems with the conditions of epistemic sovereignty. They are not incremental policy suggestions but categorical requirements derived from first principles.

\begin{enumerate}
  \item Implement epistemic infrastructure as public logic commons: adversarial interfaces, auditability by design, and traceable provenance in AI output.
  \item Mandate curriculum in formal logic, probability, Bayesian inference, and falsifiability for all civic education.
  \item Design algorithms with user-prompt transparency: no suggestion without disclosure of construction rule.
  \item Define and codify cognitive sovereignty as a right: the right to disassemble, reconstruct, and challenge all system outputs.
\end{enumerate}

\subsubsection*{IV. Formal Notation of the Autonomous Citizen}

Let $C$ denote a citizen embedded within an informational system. The predicates below formalise the necessary and sufficient conditions for epistemic autonomy, sovereignty, and civic rationalism. These are not behavioural descriptors, but logical constructs specifying the structural preconditions for a citizen to qualify as epistemically free and democratically operative.

\[
\text{Autonomous}(C) \iff \forall x\ [\text{Input}(x, C) \rightarrow (\text{Interrogate}(C, x) \wedge \text{Validate}(C, x) \wedge \neg \text{Blind\_Acceptance}(C, x))]
\]

\noindent
This definition asserts that a citizen is autonomous if and only if they apply recursive interrogation and validation to every informational input. Autonomy here is defined not by access, but by scrutiny. Passive consumption disqualifies agency.

\[
\text{Sovereign}(C) \iff \text{Epistemic\_Ownership}(C) \wedge \text{Procedural\_Agency}(C) \wedge \neg \text{Delegated\_Cognition}(C)
\]

\noindent
Sovereignty is defined as the conjunction of two core conditions: epistemic ownership (the ability to generate and evaluate knowledge independently) and procedural agency (the capacity to act based on structured rational processes). Absence of these leads to cognitive dependency and system-mediated heteronomy.

\[
\text{Civic\_Rationalism}(C) \Rightarrow \text{System\_Resistance}(C) \wedge \text{Recursive\_Cognition}(C) \wedge \text{Adversarial\_Capacity}(C)
\]

\noindent
Civic rationalism entails a structural posture of resistance against pre-structured epistemic environments. The citizen must possess recursive cognition (the ability to reason about their own reasoning processes) and adversarial capacity (the skill to test, refute, and reconstruct claims, including those presented by algorithmic systems).

\[
\text{Inert}(C) \iff \exists x\ [\text{Input}(x, C) \wedge \neg \text{Interrogate}(C, x) \wedge \text{Accept}(C, x)]
\]

\noindent
This defines epistemic inertia: the passive absorption of information without interrogation. Such a state is logically incompatible with both autonomy and sovereignty.

\[
\text{Democratic\_Validity}(C) \iff \text{Autonomous}(C) \wedge \text{Sovereign}(C) \wedge \text{Civic\_Rationalism}(C)
\]

\noindent
Finally, the model asserts that democratic validity of citizenship presupposes the co-satisfaction of all previous conditions. A subject who is merely enfranchised but procedurally inert cannot sustain the epistemic demands of a functional democracy. The informational citizen must be a logical actor within a recursive system of claims, evidence, and validation.

\subsubsection*{V. Integration into the Broader Thesis}

This subsection consolidates the logical architecture articulated above and embeds it within the central thesis: that epistemic sovereignty is not an abstract ideal, but a procedural necessity for civic legitimacy in algorithmically mediated societies.

The axiomatic foundation establishes rational apprehension and interpretive autonomy as non-negotiable preconditions for epistemic legitimacy. These axioms are then expanded into formalised propositions, providing a deductive scaffold that bridges normative philosophy with actionable political theory. Each proposition delineates a logical constraint on legitimate system design, agent sovereignty, and infrastructure accountability.

The prescriptive layer translates these abstractions into policy and institutional reform. Civic education is reconceptualised not as rote memorisation or digital literacy, but as a curriculum rooted in logic, adversarial reasoning, and falsifiability. Algorithmic design must disclose its scaffolding—no recommendation should be opaque, no prompt structurally coercive. Sovereignty is codified not merely as freedom from interference, but as the right to interrogate and reconstruct the systems that mediate cognition.

The formal logic captures the transformation from epistemic passivity to civic rationalism. It expresses, with precision, the structural conditions under which a citizen can be said to reason, dissent, and decide without mediation by unseen algorithms or inherited epistemic authority. The citizen is defined not by data access or voting capacity, but by recursive cognition, procedural agency, and resistance to automated suggestion.

In total, this model of epistemic sovereignty offers a systemic inversion of current informational hierarchies. It demands not behavioural nudges or content moderation, but foundational redesigns of how knowledge is structured, accessed, and contested. It fuses political rationality, computational architecture, and civic ethics into a schema that renders democratic agency logically defensible and structurally reproducible.

Without this procedural core, no amount of interface reform or educational access will suffice. In the absence of recursive cognitive agency, democracy degrades into a ritual of consent without comprehension, a theatre of participation devoid of epistemic independence. The theory articulated here positions logic, not intuition; transparency, not fluency; and sovereignty, not satisfaction, as the minimal conditions of political modernity.

\section{Conclusion}

Cognitive castes are no longer a speculative projection; they are an emergent political reality structured by access, capacity, and epistemic control. This paper has demonstrated that algorithmic mediation does not merely distribute information differently—it reorganises the very possibility of understanding. The architecture of knowledge, once contingent on deliberative institutions and shared referentiality, is now increasingly monopolised by cognitive elites who manipulate systems not only through data access, but through recursive comprehension and symbolic abstraction. The consequence is a political economy of intelligence stratified not by birth or capital alone, but by one’s capacity for adversarial epistemology.

To frame the trajectory of this transformation, we can envision a future scenario matrix comprised of four dominant models:

\begin{itemize}
  \item \textbf{Liberal AI:} Maximal access, minimal guidance. Individuals are flooded with information but unaided in discerning its reliability. Epistemic overload and nihilistic disengagement proliferate.
  \item \textbf{Managed AI:} Behavioural nudging cloaked in choice. Interfaces optimise for stability rather than truth, reinforcing pacification over autonomy.
  \item \textbf{Censored AI:} Rigid epistemic architectures, with top-down truth regimes codified into algorithmic constraints. Rational dissent becomes technically impossible.
  \item \textbf{Sovereign AI:} Designed for procedural transparency, auditability, and adversarial literacy. Citizens do not merely consume content—they interrogate, deconstruct, and reconstruct it.
\end{itemize}

The argument advanced here insists that emancipation is not a function of open data, decentralised platforms, or even participatory interfaces. It lies instead in the reclamation of rational autonomy as a civic and epistemic norm. To be free is not simply to choose; it is to choose with cause, to reject with reason, and to interrogate without permission.

The task ahead is monumental. It requires the construction of adversarial infrastructures, the codification of cognitive sovereignty as a civic right, and the pedagogical inversion of userhood into authorship. But no society built on interpretive outsourcing and behavioural sedation can remain politically coherent. The future of democracy hinges not on regulating content, but on constructing citizens capable of defeating systems designed to think for them.

\newpage
\printbibliography

\end{document}